%% file: main.tex
\documentclass[fleqn,usenatbib]{mnras}

\usepackage[T1]{fontenc}

\DeclareRobustCommand{\VAN}[3]{#2}
\let\VANthebibliography\thebibliography
\def\thebibliography{\DeclareRobustCommand{\VAN}[3]{##3}\VANthebibliography}

\usepackage{import}
\usepackage{lipsum}
\usepackage{bm}
\usepackage{dsfont}
\usepackage{color}
\usepackage{graphicx}	
\graphicspath{figures/}
\usepackage[tbtags]{amsmath}	
\usepackage{physics}
\usepackage{xcolor}

\newcommand{\HI}{\ion{H}{I}}
\newcommand{\HII}{\ion{H}{II}}
\renewcommand{\*}[1]{\bm{#1}}

\title[Inference from 21-cm $P(k)$ through MNRE]{Constraining the X-ray heating and reionization using 21-cm power spectra with Marginal Neural Ratio Estimation}

\author[Saxena et al.]{Anchal Saxena,$^{1}$\thanks{E-mail: a.saxena@rug.nl}
Alex Cole,$^{2}$
Simon Gazagnes,$^{3}$
P.\ Daniel Meerburg,$^{1}$
Christoph Weniger,$^{2}$
 \newauthor  
and Samuel J. Witte$^{2, 4}$
\\
$^{1}$Van Swinderen Institute, University of Groningen, Nijenborgh 4, 9747 AG Groningen, The Netherlands\\
$^{2}$Gravitation Astroparticle Physics Amsterdam (GRAPPA), Institute for Theoretical Physics Amsterdam and \\Delta Institute for Theoretical Physics, University of Amsterdam, Science Park 904, 1098 XH Amsterdam, The Netherlands\\
$^{3}$Department of Astronomy, The University of Texas at Austin, 2515 Speedway, Stop C1400, Austin, TX 78712-1205, USA \\
$^{4}$Departament de F\'{i}sica Qu\`{a}ntica i Astrof\'{i}sica and Institut de Ciencies del Cosmos \\
Universitat de Barcelona, Diagonal 647, E-08028 Barcelona, Spain
}

\date{Accepted XXX. Received YYY; in original form ZZZ}

\pubyear{2022}

\begin{document}
\label{firstpage}
\pagerange{\pageref{firstpage}--\pageref{lastpage}}
\maketitle

\begin{abstract}
Cosmic Dawn (CD) and Epoch of Reionization (EoR) are epochs of the universe which host invaluable information about the cosmology and astrophysics of X-ray heating and hydrogen reionization. Radio interferometric observations of the 21-cm line at high redshifts have the potential to revolutionize our understanding of the universe during this time. However, modeling the evolution of these epochs is particularly challenging due to the complex interplay of many physical processes. This makes it difficult to perform the conventional statistical analysis using the likelihood-based Markov-Chain Monte Carlo ({\small MCMC}) methods, which scales poorly with the dimensionality of the parameter space. In this paper, we show how the Simulation-Based Inference (SBI) through Marginal Neural Ratio Estimation ({\small MNRE}) provides a step towards evading these issues. We use \texttt{21cmFAST} to model the 21-cm power spectrum during CD-EoR with a six-dimensional parameter space. With the expected thermal noise from the Square Kilometre Array (SKA), we are able to accurately recover the posterior distribution for the parameters of our model at a significantly lower computational cost than the conventional likelihood-based methods. We further show how the same training dataset can be utilized to investigate the sensitivity of the model parameters over different redshifts. Our results support that such efficient and scalable inference techniques enable us to significantly extend the modeling complexity beyond what is currently achievable with conventional {\small MCMC} methods.
\end{abstract}




\section{Introduction}
\label{sec:intro}
\import{sections/}{introduction.tex}

\section[Implementation of MNRE using swyft]{Implementation of MNRE using \lowercase{\textit{swyft}}}
\label{sec:framework}
\import{sections/}{swyft_framework.tex}

\section{Simulations and training data}
\label{sec:sims}
\import{sections/}{sims_data.tex}

\section{Results}
\label{sec:results}
\import{sections/}{results.tex}

\section{Summary}
\label{sec:summary}
\import{sections/}{summary.tex}

\section*{Acknowledgements}
We thank the Center for Information Technology of the University of Groningen for their support and for providing access to the Peregrine high performance computing cluster. P.D.M acknowledges support from the Netherlands organization for scientific research (NWO) VIDI grant (dossier 639.042.730). S.G acknowledges support from the Harlan J. Smith McDonald fellowship. S.J.W. and C.W. are supported by the European Research Council (ERC) under the European Union’s Horizon 2020 research and innovation programme (Grant agreement No. 864035 - Undark).  The project has been partially funded by the Netherlands eScience Center, grant number ETEC.2019.018.

\section*{Data Availability}
Accompanying code is available at \url{https://github.com/anchal-009/swyft21cm}. The data underlying this article will be shared on reasonable request to the corresponding author.



\bibliographystyle{mnras}
\bibliography{references} 



\appendix
\section{Coverage of the network}
\begin{figure*}
    \centering
    \includegraphics[width=0.98\linewidth]{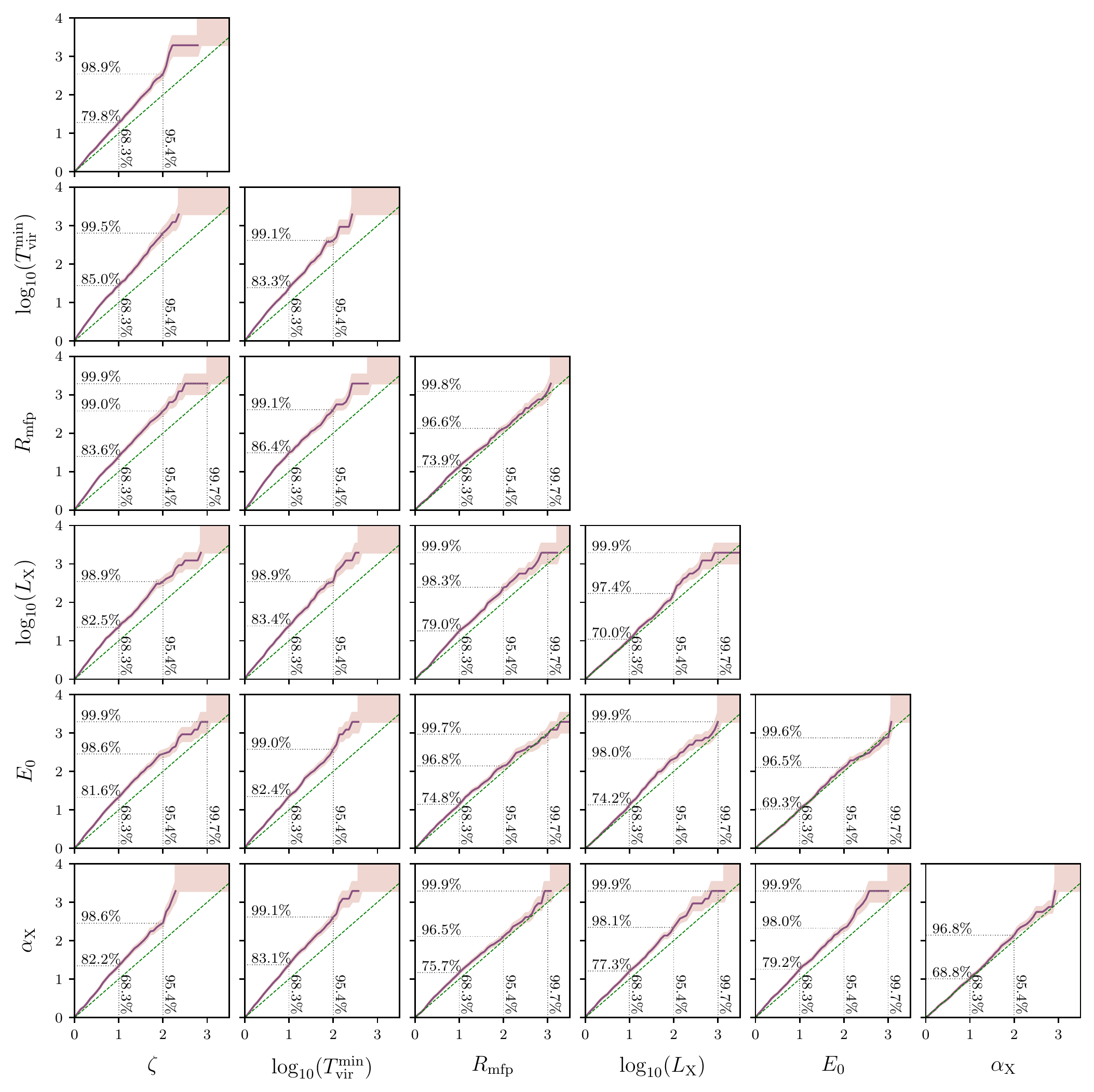}
    \caption{Empirical expected coverage probability ($1-\hat{\alpha}$) of the trained network as a function of confidence level ($1-\alpha$) for all 1D and 2D marginal posteriors. In case of a perfect coverage, the purple line coincides with the green dashed line.}
    \label{fig:coverage}
\end{figure*}
With the trained network $d_{\phi} (\*x, \*\theta)$, we can quickly estimate the posterior $p(\*\theta|\*x)$ for any mock observation $\*x$. This allows us to test the statistical properties of the Bayesian inference. We evaluate the nominal and empirical expected coverage probabilities to check the consistency of the trained network. Given a set of $n$ i.i.d.\ samples ($\*x_i, \*\theta_i^*$) $\sim p(\*x, \*\theta)$, the empirical expected coverage probability of the (1 - $\alpha$) highest posterior density regions (HPDR) for the posterior estimator $\hat{p}(\*\theta|\*x)$ is given as \citep{https://doi.org/10.48550/arxiv.2110.06581}
\begin{equation}
    1 - \hat{\alpha} = \frac{1}{n} \sum_{i=1}^{n} \mathds{1}\left[\*\theta_i^* \in \Theta_{\hat{p}(\*\theta|\*x_i)}(1-\alpha)\right]\,,
\end{equation}
where $\Theta_{\hat{p}(\*\theta|\*x_i)}(1-\alpha)$ function gives the ($1-\alpha$) HPDR of $\hat{p}(\*\theta|\*x)$ for the mock data $\*x_i$ with the ground truth $\*\theta_i^*$. We then compare it with the nominal expected coverage probability, which is equal to the confidence level ($1 - \alpha$). For an estimator with perfect coverage, the empirical coverage probability is equal to the nominal coverage probability, so when we randomly generate $n$ samples ($\*x_i, \*\theta_i^*$) $\sim p(\*x, \*\theta)$, the ground truth $\*\theta_i^*$ lies outside the ($1-\alpha$) HPDR in $\alpha$ of the cases \citep{Cole_2022}. 

We re-parameterize $\alpha$ ($\hat{\alpha}$) in terms of a new variable $z$ which is $1 - \alpha/2$ ($1 - \hat{\alpha}/2$) quantile of the standard normal distribution. This means that the $(1, 2, 3)\sigma$ regions correspond to $z=(1, 2, 3)$ with ($1-\alpha$) = (0.6827, 0.9545, 0.9997). The uncertainties on the empirical expected coverage probability follow from the finite number of samples ($n$) and are estimated using the Jeffreys interval \citep{Cole_2022}. In Figure~\ref{fig:coverage}, we show the empirical expected coverage probability of the network as a function of confidence levels for all 1D and 2D marginal posteriors. We find that in all cases, they match to good precision.

\section{Comparison with 21CMMC}
\begin{figure}
    \centering
    \includegraphics[width=\linewidth]{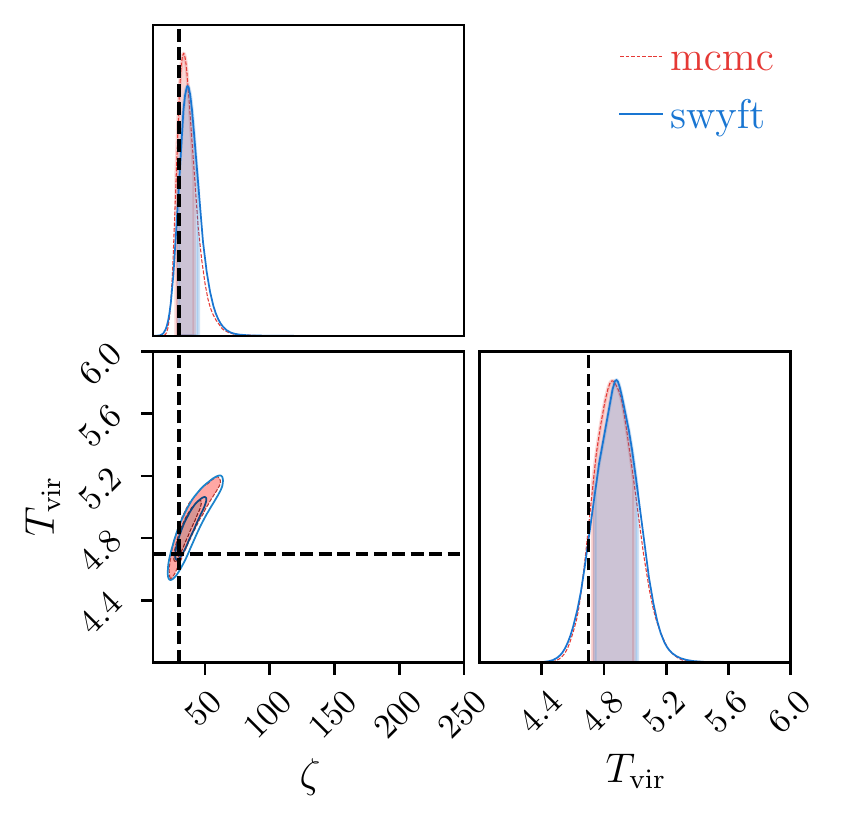}
    \caption{Comparison of the 1D and 2D marginal posteriors estimated from 21CMMC (red) with MNRE (blue) for the FAINT GALAXIES reionization model $(\zeta, \log_{10}(T_{\rm vir}^{\rm min}))$ = (30, 4.70).}
    \label{fig:mcmc_swyft}
\end{figure}
In this section, we compare the posteriors obtained from {\small MNRE} with an {\small MCMC} sampling-based method, {\small 21CMMC}. As sampling our joint six-dimensional parameter space with the likelihood-based approach is computationally very demanding, we shrink our parameter space to two dimensions, including $\zeta$ and $\log_{10}(T_{\rm vir}^{\rm min})$. We keep the other parameters fixed to $\{R_{\rm mfp}, \log_{10}(L_{\rm X}), E_0, \alpha_{\rm X}\}$ = \{15, 40.5, 0.5, 1\} and target redshifts $z$ = 10, 9 and 8. 

To set up {\small 21CMMC}, we use 48 random walkers with 2000 iterations each, generating $\sim 10^5$ samples. On the other hand, the training data for {\small MNRE} consists of $10^4$ simulations. The mock observation with $(\zeta, \log_{10}(T_{\rm vir}^{\rm min}))$ = (30, 4.70) is generated using a different realization of the density field from the one used in sampling. In Figure~\ref{fig:mcmc_swyft}, we show the posteriors obtained from {\small 21CMMC} (red) and {\small MNRE} (blue). The 1D and 2D marginal posteriors obtained from {\small MNRE} are in good agreement with {\small 21CMMC} at a significantly reduced computational cost. These results are consistent with the findings of \citet{Zhao_2022}.

\section{Impact of including modeling uncertainty}
\label{app:mod_uncert}
\begin{figure*}
    \centering
    \includegraphics[width=\linewidth]{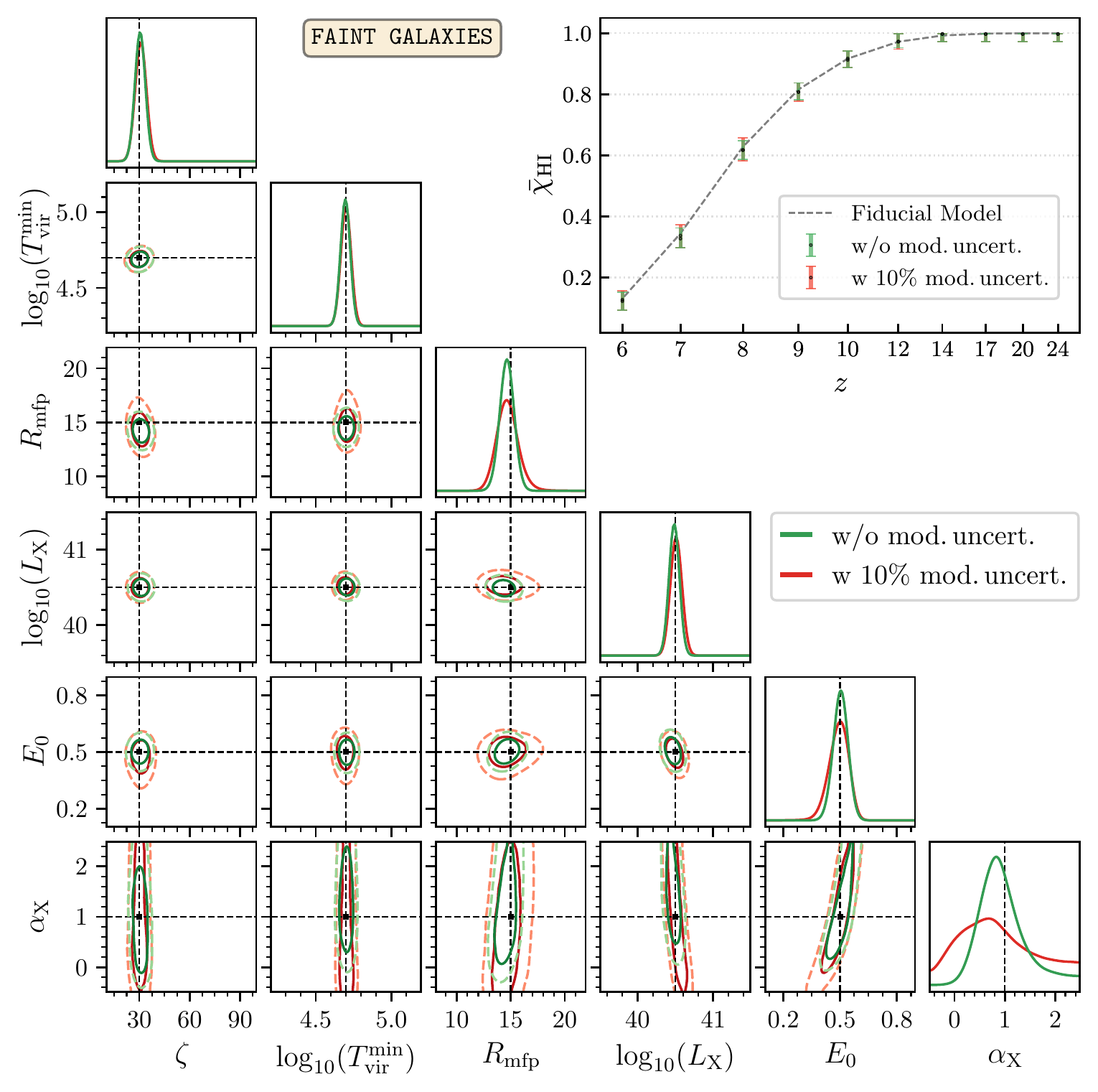}
    \caption{Recovered 1D and 2D marginals with (red) and without (green) including 10\% modeling uncertainty on the 21-cm power spectra. The dashed lines denote the input parameters \{$\zeta, \log_{10}(T_{\rm vir}^{\rm min}), R_{\rm mfp}, \log_{10}(L_{\rm X}), E_0, \alpha_{\rm X}\}$ = \{30, 4.70, 15, 40.5, 0.5, 1\}. The inset plot shows the recovered reionization history.}
    \label{fig:postModUncert}
\end{figure*}

In this section, in addition to thermal noise, we consider an additional source of uncertainty due to the EoR modeling. We assume a constant multiplicative error of 10\% to take into account the errors in semi-numerical approximations. This is added in quadrature to the thermal noise uncertainty to get the 21-cm power spectrum uncertainty
\begin{equation}
    \label{eq:incl_modUncert}
    \sigma(k_i, z) = \sqrt{\sigma_{\rm therm}^2 (k_i, z) + \sigma_{\rm mod}^2 (k_i, z)}\,.
\end{equation}
In Figure~\ref{fig:postModUncert}, we show the recovered 1D and 2D marginal posteriors assuming a 10\% modeling uncertainty (red), and compare it with the constraints derived by excluding this error (green). We find that including the modeling uncertainty results in wider posteriors, which is consistent with \citet{Greig_2015}. The inferred model parameters and the corresponding 16th and 84th percentiles for both scenarios are tabulated in Table~\ref{tab:paramsRecovModUncert}. We note that this analysis does not require re-running any 21-cm signal simulations. The existing training data with the modified noise model given by equation~(\ref{eq:incl_modUncert}), which is sampled on-the-fly during the training of the network, can be re-used. This example demonstrates the flexibility and efficiency of our approach.

\begin{table*}
 \renewcommand{\arraystretch}{1.6}
 \centering
 \caption{The inferred parameter values and the associated 16th and 84th percentiles for the posteriors shown in Figure~\ref{fig:postModUncert} with and without including 10\% modeling uncertainty.}
 \label{tab:paramsRecovModUncert}
 \begin{tabular}{lllllll}
  \hline
  Model & $\zeta$ & $\log_{10}(T_{\rm vir}^{\rm min})$ & $R_{\rm mfp}$ & $\log_{10}(L_{\rm X})$ & $E_0$ & $\alpha_{\rm X}$\\
  \hline
  w/o 10\% mod.\ uncert. & $30.25_{-1.80}^{+2.70}$ & $4.70_{-0.02}^{+0.03}$ & $14.65_{-0.56}^{+0.56}$ & $40.49_{-0.06}^{+0.04}$ & $0.50_{-0.03}^{+0.03}$ & $0.84_{-0.39}^{+0.39}$\\
  w 10\% mod.\ uncert. & $31.15_{-2.70}^{+2.70}$ & $4.70_{-0.03}^{+0.02}$ & $14.65_{-0.84}^{+0.84}$ & $40.51_{-0.06}^{+0.06}$ & $0.50_{-0.05}^{+0.04}$ & $0.69_{-0.78}^{+0.69}$\\
  \hline
 \end{tabular}
\end{table*}

\section{Impact of the size of training set on the posteriors}
\label{app:trainSubset}
\begin{table*}
 \renewcommand{\arraystretch}{1.6}
 \centering
 \caption{The inferred parameter values and the associated 16th and 84th percentiles for the posteriors shown in Figure~\ref{fig:postSubset}.}
 \label{tab:paramsRecovSubset}
 \begin{tabular}{lllllll}
  \hline
  Model & $\zeta$ & $\log_{10}(T_{\rm vir}^{\rm min})$ & $R_{\rm mfp}$ & $\log_{10}(L_{\rm X})$ & $E_0$ & $\alpha_{\rm X}$\\
  \hline
  $n_{\rm samp} = 2 \times 10^4$ & $30.25_{-1.80}^{+2.70}$ & $4.70_{-0.02}^{+0.03}$ & $14.65_{-0.56}^{+0.56}$ & $40.49_{-0.06}^{+0.04}$ & $0.50_{-0.03}^{+0.03}$ & $0.84_{-0.39}^{+0.39}$\\
  $n_{\rm samp} = 10^4$ & $30.25_{-2.70}^{+2.70}$ & $4.70_{-0.03}^{+0.02}$ & $14.37_{-0.56}^{+0.56}$ & $40.49_{-0.04}^{+0.06}$ & $0.50_{-0.04}^{+0.03}$ & $0.81_{-0.36}^{+0.36}$\\
  \hline
 \end{tabular}
\end{table*}
In this section, we investigate the size of the training set needed to achieve the convergence for {\small MNRE}. Our default training set contains $2\times10^4$ ($n_{\rm samp}$) 21-cm power spectra samples. We re-train the neural ratio estimator with a subset of training set with $n_{\rm samp} = 10^4$ to estimate the posterior distribution of model parameters. 

\begin{figure*}
    \centering
    \includegraphics[width=\linewidth]{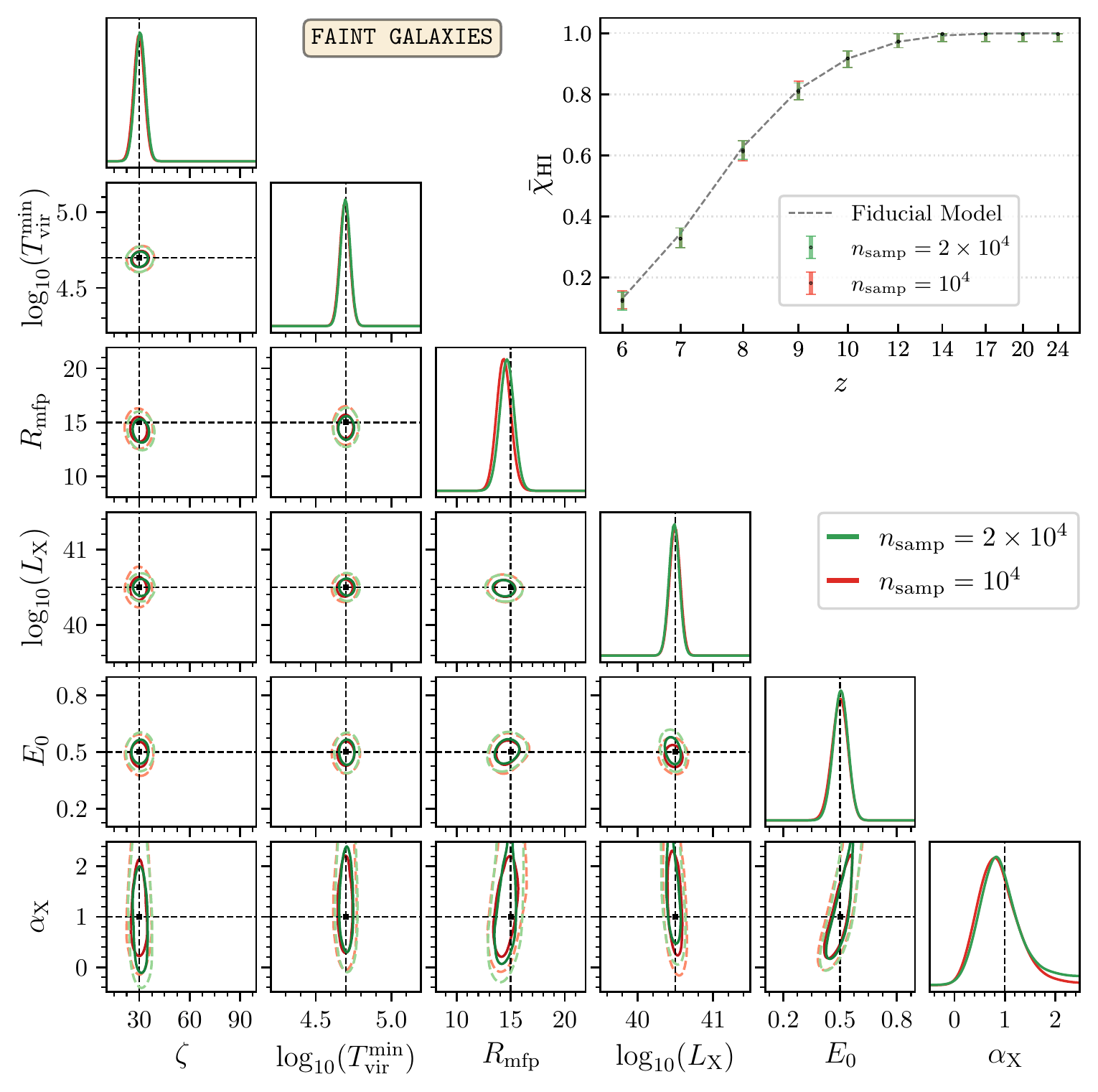}
    \caption{Recovered 1D and 2D marginals with the number of samples in the training data $n_{\rm samp} = 2\times 10^4$ (green) and $n_{\rm samp} = 10^4$ (red). The dashed lines denote the input parameters \{$\zeta, \log_{10}(T_{\rm vir}^{\rm min}), R_{\rm mfp}, \log_{10}(L_{\rm X}), E_0, \alpha_{\rm X}\}$ = \{30, 4.70, 15, 40.5, 0.5, 1\}. The inset plot shows the recovered reionization history.}
    \label{fig:postSubset}
\end{figure*}

In Figure~\ref{fig:postSubset}, we present and compare the recovered 1D and 2D marginal posteriors generated from $n_{\rm samp} = 2\times10^4$ (green) and $10^4$ (red) samples. The inset plot shows the 2$\sigma$ constraints on reionization history. The inferred model parameters and the corresponding 16th and 84th percentiles for both scenarios are tabulated in Table~\ref{tab:paramsRecovSubset}. The posteriors on model parameters for both cases match to excellent precision, which indicates the convergence of {\small MNRE}. Therefore, $\sim 10^4$ simulations are sufficient to preserve accuracy in our {\small SBI} framework which makes it $3 - 10$ times more computationally efficient than the classical methods of inference. 


\bsp	
\label{lastpage}
\end{document}

%% file: sections/introduction.tex
The Cosmic Dawn (CD) marks the formation of the first sources of light, which produced high-energy X-ray and UV radiation. The radiation from these sources heated up the intergalactic medium (IGM) and initiated the Epoch of Reionization (EoR), during which the IGM transitioned from a neutral to ionized state \citep{2001PhR...349..125B,2006PhR...433..181F,Pritchard_2012}. The astrophysics driving the heating and reionization process is still poorly understood, with large uncertainties on the properties of the sources which dominantly contributed to these epochs (e.g., their star formation efficiency, ionizing efficiency, and their X-ray luminosity). Observations of the high redshift quasar spectra \citep{2001AJ....122.2850B, 2003AJ....125.1649F, boera19}, electron scattering optical depth from the Cosmic Microwave Background (CMB) \citep{2003ApJ...583...24K,2011ApJS..192...18K,2020A&A...641A...6P}, and the luminosity function and clustering properties of Lyman-$\alpha$ emitters \citep{Jensen_2012, Dijkstra_2014, Bouwens_2016, Gangolli_2020} currently provide some constraints on the astrophysical evolution of the CD and EoR. The 21-cm line associated with the spin-flip hyperfine transition of the hydrogen atom offers the most promising probe to study these eras.

There are a large number of ongoing radio interferometric experiments, including GMRT \citep{2013MNRAS.433..639P}, HERA \citep{DeBoer_2017}, LOFAR \citep{mertens20, 2020MNRAS.493.4728G}, LWA \citep{Eastwood_2019}, MWA \citep{barry19, li19} and PAPER \citep{kolopanis19}. These experiments target the detection of the 21-cm signal by quantifying its spatial fluctuations using various Fourier statistics. We get increasingly interesting upper limits on the 21-cm power spectra from these experiments, some of which already enable us to rule out certain astrophysical models \citep{2020MNRAS.493.4728G, 2021MNRAS.503.4551G, 2020MNRAS.498.4178M, 2021MNRAS.501....1G, Abdurashidova_2022, https://doi.org/10.48550/arxiv.2210.04912}. The upcoming SKA \citep{2015aska.confE...1K,mellema2015hi} is expected to detect the 21-cm power spectrum and owing to its high sensitivity, it is likely that SKA will also be able to do the full tomography of the 21-cm signal.

Once the signal is detected, the next goal would be to constrain the parameters of the CD-EoR models to pin down the astrophysics of the early universe. Modeling the 21-cm signal from CD-EoR using full radiative transfer simulations \citep{Mellema_2006, Ghara_2015} is computationally expensive and unfeasible to perform parameter inference. To overcome this challenge, various approximate and efficient semi-numerical models are used to model the signal accurately at scales $\geq$ 1 Mpc \citep{Zahn_2011}. The traditional framework that is used to explore parameter space is {\small 21CMMC}\footnote{\url{https://github.com/BradGreig/21cmMC}} \citep{Greig_2015,Greig_2017,Greig_2018,Park_2019}, which uses a semi-numerical framework of the 21-cm signal simulator \texttt{21cmFAST}\footnote{\url{https://github.com/21-cmfast/21cmFAST}} \citep{Mesinger_2010} and embed this code in a Markov-Chain Monte Carlo ({\small MCMC}) sampler. While {\small 21CMMC} is quite powerful in systematically performing parameter inference, it becomes computationally quite expensive once we take into account the inhomogeneous X-ray heating in the simulations. Alternatively, one can use analytical models of the 21-cm signal during CD-EoR \citep{Qin_2022, https://doi.org/10.48550/arxiv.2302.08506}. Quite generally, as the dimensionality of parameter space increases, it takes longer for an {\small MCMC}, which samples the full joint posterior, to converge.

To circumnavigate these problems, machine learning techniques have been explored in various astrophysical and cosmological problems. In the context of 21-cm cosmology, one common approach is to use emulators, which are trained using artificial neural networks to replace actual simulations. This makes the likelihood evaluations and, consequently, the parameter inference significantly faster \citep{Shimabukuro_2017, Kern_2017, Schmit_2017, Tiwari_2022}. However, the application of emulators is currently limited to low-order summary statistics. The likelihood could become intractable for higher-order information such as the full 3D 21-cm images. For a tractable likelihood function, the traditional {\small MCMC} algorithm can be used to sample from the posterior distribution. However, when the likelihood itself is intractable, techniques such as the Approximate Bayesian Computation ({\small ABC}) \citep{Toni_2008} can be used to sample from the approximate posterior. This approach uses simulated datasets to avoid the likelihood evaluations; however, it requires the introduction of summary statistics, which can significantly affect the quality of the approximation.

These issues can be resolved by performing a Simulation-Based Inference ({\small SBI}) \citep{doi:10.1073/pnas.1912789117, pmlr-v89-papamakarios19a, Alsing_2019}, where deep learning algorithms along with the {\small ABC} are used to estimate the posterior distribution. In this work, we will apply the Marginal Neural Ratio Estimation ({\small MNRE}) algorithm \citep{Miller:2021hys} using \textit{swyft}\footnote{\url{https://github.com/undark-lab/swyft}}\citep{Miller:2022shs}. It directly estimates the marginal likelihood-to-evidence ratios through neural networks, which makes it much more efficient than sampling the full joint posterior with an {\small MCMC}. In addition, {\small MNRE} offers the flexibility to ignore large numbers of nuisance parameters, learning only the parameters of interest. This has already been applied for the cosmological parameter inference from the CMB power spectra \citep{Cole_2022}, reconstructing the halo clustering and halo mass function from N-body simulations \citep{https://doi.org/10.48550/arxiv.2206.11312}, and gravitational lensing analyses \citep{https://doi.org/10.48550/arxiv.2209.09918}.

In this work, we use this framework for the astrophysical parameter inference with the 21-cm power spectrum from the CD-EoR. In a recent study, \citet{Zhao_2022} have performed the reionization parameter inference from the EoR using the density estimation likelihood free inference ({\small DELFI}). Their analysis, however, was limited to a two-dimensional parameter space to model the 21-cm signal during the EoR. Here, we extend the parameter space to six dimensions to also include the parameters that govern the inhomogeneous X-ray heating during the CD. In this case, a single 21-cm power spectrum simulation is $\sim 5$ times slower than the former 2D parameter space. This implies that an MCMC for the six-dimensional parameter space would be even slower because of the typical exponential scaling of the required samples as a function of the number of parameters. Moreover, such analysis with conventional methods while also co-varying the cosmic seed for the forward models is pushed out of the realm of feasibility.

Generating the simulated dataset and performing the inference with {\small MNRE} are two independent processes within \textit{swyft}. This allows us to utilize the same training dataset for various applications. To highlight this aspect of \textit{swyft}, in a worked-out example, we will let the neural network determine which set of parameters are sensitive at which redshifts at no extra cost of 21-cm simulations. The distribution of integration time over different redshifts can be considered a proxy to determine which part of the data each parameter is most sensitive to. This could be indicative of the possible degeneracies between parameters for more complex astrophysical models of the 21-cm signal.

This paper is organized as follows. In section~\ref{sec:framework}, we briefly outline the implementation of {\small MNRE} using \textit{swyft}. In section~\ref{sec:sims}, we describe the 21-cm signal modeling and the parameters of interest. In section~\ref{sec:results}, we present the posterior inference and investigate the sensitivity of model parameters in different redshift ranges. We conclude in section~\ref{sec:summary}. Throughout this work, we assumed a $\Lambda$CDM universe with cosmological parameters $\Omega_{\rm m} = 0.308$, $\Omega_{\rm b} = 0.048$, $\Omega_{\Lambda} = 0.692$, $h=0.678$ and $\sigma_8 = 0.81$ \citep{2016A&A...594A..13P}.

%% file: sections/swyft_framework.tex
The probability distribution of model parameters $\*\theta$ for a given observation $\*x$ follows from Bayes' theorem
\begin{equation}
    \label{eq:bayes}
    p(\*\theta|\*x) = \frac{p(\*x|\*\theta)}{p(\*x)}\, p(\*\theta)\,,
\end{equation}
where $p(\*x|\*\theta)$ is the likelihood of the data $\*x$ for given parameters $\*\theta$, $p(\*\theta)$ is the prior probability distribution over the parameters and $p(\*x)$ is the evidence of the data. 

In {\small SBI}, the information about the likelihood is implicitly accessed via a stochastic simulator, which maps from input parameters $\*\theta$ to data $\*x$. We generate sample-parameter pairs from this simulator $\{(\*x^1, \*\theta^1)\, (\*x^2, \*\theta^2), \cdots\}$. Here $\*\theta^i$ is typically drawn from the prior, so these pairs are drawn from the joint distribution $p(\*x, \*\theta)$. These pairs are used to train a neural network to approximate the likelihood-to-evidence ratio, a procedure known as Neural Ratio Estimation ({\small NRE}) \citep{hermans2020likelihoodfree, pmlr-v119-hermans20a, https://doi.org/10.48550/arxiv.2002.03712}. Following equation~(\ref{eq:bayes}), this ratio (which we denote $r(\*x, \*\theta)$) can be expressed as
\begin{equation}
    r(\*x, \*\theta) \equiv \frac{p(\*x|\*\theta)}{p(\*x)} = \frac{p(\*\theta|\*x)}{p(\*\theta)} = \frac{p(\*x, \*\theta)}{p(\*x)\,p(\*\theta)}\,.
\end{equation}
In other words, $r(\*x, \*\theta)$ is equal to the ratio of the joint probability density $p(\*x, \*\theta)$ to the product of marginal probability densities $p(\*x)\, p(\*\theta)$.
A binary classifier $d_{\*\phi}(\*x, \*\theta)$ is then trained to distinguish between jointly-drawn and marginally-drawn pairs. Here $\*\phi$ denotes the learnable parameters of the model, which are updated as the model is trained.

More precisely, we introduce a binary label $y$ to denote whether a pair was drawn jointly ($y=1$) or marginally ($y=0$). Strictly speaking, $y$ is a random variable.
The output of the classifier (assuming it is trained well) approximates the probability that a sample-parameter pair ($\*x, \*\theta$) is drawn jointly ($y=1$), i.e., \
\begin{equation*}
    \begin{split}
        d_{\*\phi} (\*x, \*\theta) &\approx p(y=1|\*x, \*\theta) \\
        &= \frac{p(\*x, \*\theta|y=1) p(y=1)}{p(\*x, \*\theta|y=1) p(y=1) + p(\*x, \*\theta|y=0) p(y=0)} \\
        &= \frac{p(\*x, \*\theta)}{p(\*x, \*\theta) + p(\*x) p(\*\theta)}\,,
    \end{split}
\end{equation*}
where we assumed $p(y=0) = p(y=1) = \frac{1}{2}$. This learning problem is associated with a binary cross-entropy loss function
\begin{equation}
    - \int \left[p(\*x, \*\theta) \ln d_\phi (\*x, \*\theta) + p(\*x)\, p(\*\theta) \ln \{1 - d_\phi (\*x, \*\theta)\}\right] \dd{\*x} \dd{\*\theta}\,,
\end{equation}
which is minimized using stochastic gradient descent to find the optimal parameters $\*\phi$ of the network. The binary classifier is simply a dense neural network with a few hidden layers. Once the network is trained, it results in 
\begin{equation}
    d_{\*\phi} (\*x, \*\theta) \approx \frac{p(\*x, \*\theta)}{p(\*x, \*\theta) + p(\*x) p(\*\theta)} = \frac{r(\*x, \*\theta)}{r(\*x, \*\theta) + 1}\,,
\end{equation}
which can be re-written as 
\begin{equation}
        r(\*x, \*\theta) \approx \frac{d_{\*\phi}(\*x, \*\theta)}{d_{\*\phi}(\*x, \*\theta)-1} \implies
        p(\*\theta|\*x) \approx \frac{d_{\*\phi}(\*x, \*\theta)}{d_{\*\phi}(\*x, \*\theta)-1}\,p(\*\theta)
\end{equation}
to estimate the posterior probability distribution. This procedure can directly estimate marginal posteriors by omitting model parameters from the network's input, a variant called Marginal Neural Ratio Estimation ({\small MNRE}). In this work, we use {\small MNRE} as implemented in the software package \emph{swyft} \citep{Miller:2022shs}.

%% file: sections/sims_data.tex
\subsection{21cmFAST}
To model the 21-cm signal and the underlying astrophysics of heating and reionization, we use the publicly available semi-numerical formalism, \texttt{21cmFAST} \citep{Mesinger_2010}. We first generate the initial density perturbation at $z = 300$ on a high-resolution $1024^3$ grid. These perturbations are evolved using the Zel'dovich approximation \citep{1970A&A.....5...84Z} at later redshifts. To produce the ionization map, the high-resolution density field is first mapped on a coarser grid. Then, \texttt{21cmFAST} uses an excursion-set based formalism \citep{Furlanetto_2004} to identify the ionized regions by comparing the number of ionizing photons with the number of baryons within the spheres of decreasing radius $R_{\rm min}\leq R\leq R_{\rm max}$. Here, $R_{\rm min}$ depends on the spatial resolution of the simulation, and $R_{\rm max}$ is the maximum horizon for ionizing photons (see Section~\ref{subsub: mfp}). A grid point located at ($\mathbf{x}, z$) is considered fully ionized if for any $R_{\rm min} \leq R \leq R_{\rm max}$
\begin{equation}
    \label{eq:ionz_condition}
    \zeta f_{\rm coll} (\mathbf{x}, z, R, M_{\rm min}) \geq 1\,,
\end{equation}
where $\zeta$ represents the ionizing efficiency (see Section~\ref{subsub: ionz_eff}) and $f_{\rm coll} (\mathbf{x}, z, R, M_{\rm min})$ is the fraction of collapsed matter within a spherical region of radius $R$ centered at $(\mathbf{x}, z)$, which depends on the minimum mass of the halo formation $M_{\rm min}$ \citep{1974ApJ...187..425P, 1999MNRAS.308..119S}. The cells that do not satisfy Equation~(\ref{eq:ionz_condition}) are assigned a partial ionization fraction, $\zeta f_{\rm coll} (\mathbf{x}, z, R_{\rm min})$. The resulting ionization map is then converted into the 21-cm brightness temperature map using \citep{Furlanetto_2006}
\begin{equation}
    \begin{aligned}
        \delta T_{\rm b} &= 27 (1 - x_{\HII}) \left( 1+\delta_b \right) \left( \frac{\Omega_b h^2}{0.023} \right) \left( \frac{0.15}{\Omega_m h^2} \frac{1+z}{10} \right)^{1/2} \\ &\times \left( \frac{T_{\rm S} - T_\textrm{CMB}}{T_\textrm{CMB}} \right) \left[\frac{\partial_r v_r}{(1+z)H(z)}\right]\,,
    \end{aligned}
\end{equation}
where $x_{\HII}$ is the ionization fraction, $\delta_b$ is the baryon overdensity, $\Omega_m$ is the matter density, $\Omega_b$ is the baryon density, $h$ is the Hubble parameter, $T_{\rm S}$ and $T_\mathrm{CMB}$ are the spin temperature and CMB temperature respectively, and the last term takes into account the velocity gradient along the line of sight.

The spin temperature $T_{\rm S}$ can couple to (i) the CMB temperature $T_{\rm CMB}$, in which case $\delta T_{\rm b} = 0$, (ii) the kinetic gas temperature $T_{\rm K}$ through collisional coupling and (iii) the Ly-$\alpha$ color temperature $T_{\rm C}$ through the Wouthuysen–Field coupling \citep{1952AJ.....57R..31W}, where $T_{\rm C} \approx T_{\rm K}$. To track the evolution of the gas temperature, \texttt{21cmFAST} simulates the inhomogeneous heating of the IGM by X-rays by integrating the angle-averaged specific X-ray emissivity ($\epsilon_{\rm X}$) along the lightcone for each cell. The specific X-ray emissivity is given as \citep{Mesinger_2010, Greig_2017}
\begin{equation}
    \epsilon_{\rm X}({\bm x}, E, z) = \frac{L_{\rm X}}{{\rm SFR}} \left[\rho_{\rm crit, 0}\Omega_{b}f_{\star}(1+\delta_{\rm nl})\frac{{\rm d}f_{\rm coll}(z)}{{\rm d}t}\right]\,,
\end{equation}
where $\rho_{\rm crit, 0}$ is the current critical density, $f_{\star}$ is the fraction of baryons in stars, $\delta_{\rm nl}$ is the evolved density. The term enclosed in square brackets is the star-formation rate (SFR) density along the lightcone. $L_{\rm X}$ is the specific X-ray luminosity which is assumed to follow a power law, $L_{\rm X} \propto E^{-\alpha_{\rm X}}$. The photons below an energy threshold $E_0$ are absorbed by the interstellar medium. The X-ray efficiency is normalized by quantifying an integrated soft-band ($<$ 2 keV) luminosity per SFR
\begin{equation}
    \frac{L_{\rm X < 2\, keV}}{\rm SFR} = \int_{E_0}^{\rm 2\, keV} \left(\frac{L_{\rm X}}{\rm SFR}\right) {\rm d}E\,.
\end{equation}

The semi-numerical model adopted in this work consists of six astrophysical parameters which govern the evolution of the 21-cm signal during the CD-EoR. We briefly describe each of these parameters and the adopted priors below.

\subsubsection{Ionizing efficiency, ($\zeta$)}
\label{subsub: ionz_eff}
The UV ionizing efficiency of high redshift galaxies can be expressed in terms of various factors as \citep{2001PhR...349..125B, Mesinger_2010}
\begin{equation}
    \zeta = 30 \left(\frac{f_{\rm esc}}{30}\right) \left(\frac{f_\star}{0.05}\right) \left(\frac{N_{\gamma/b}}{4000}\right) \left(\frac{2}{1+n_{\rm rec}}\right)\,,
\end{equation}
where $f_{\rm esc}$ is the fraction of ionizing photons that escape into the intergalactic medium (IGM), $f_\star$ is the fraction of galactic gas in stars, $N_{\gamma/b}$ is the number of ionizing photons produced per baryon in stars and $n_{\rm rec}$ is the average number of times a hydrogen atom recombines. We assume a single population of efficient star-forming galaxies (a constant ionizing efficiency for all the galaxies) hosted by haloes with a sufficient mass.

The timing and duration of reionization strongly depend on $\zeta$. Large values of $\zeta$ will speed up the ionization process if we keep the other parameters fixed. We adopt a flat prior $\zeta \in (10, 100)$, although an extended range with the upper limit of $\zeta = 250$ has also been studied in \citet{Greig_2017} to explore the models where the EoR is driven by rare, very bright galaxies.

\subsubsection{Minimum virial temperature of haloes, $T_{\rm vir}^{\rm min}$}
The minimum threshold for a halo to host a star-forming galaxy is defined in terms of its virial temperature, $T_{\rm vir}^{\rm min}$. It is related to the mass of the halo \citep{2001PhR...349..125B} as 
\begin{equation}
    M_{\rm vir}^{\rm min} = \frac{10^8}{h} \left[\frac{0.6}{\mu}\frac{10}{1 + z}\frac{T_{\rm vir}^{\rm min}}{1.98 \times 10^4}\right]^{3/2} \left[\frac{\Omega_m}{\Omega_m^z} \frac{\Delta_c}{18\pi^2}\right]^{-1/2} M_{\sun}\,,
\end{equation}
where $\mu$ is the mean molecular weight, $\Omega_m^z=\Omega_m(z)$, and $\Delta_c = 18\pi^2 + 82d - 39d^2$ where $d = \Omega_m^z - 1$. The choice of $T_{\rm vir}^{\rm min}$ determines the cut-off in the UV luminosity function. Galaxies that are hosted within a halo with $T_{\rm vir}<T_{\rm vir}^{\rm min}$ have no contribution to star formation due to internal feedback processes. We note that $T_{\rm vir}^{\rm min}$ has a significant impact on both the EoR and the Epoch of Heating (EoH) because, within the \texttt{21cmFAST} framework, the physics of star-formation drives both the X-ray heating and ionization fields.

We adopt a flat prior on $T_{\rm vir}^{\rm min} \in (10^4, 10^6)$ K. The minimum temperature required for efficient atomic cooling defines our lower limit of $T_{\rm vir}^{\rm min} = 10^4$ K, and the upper limit is consistent with the observation of Lyman break galaxies at high redshifts \citep{Kuhlen_2012, Barone_Nugent_2014}. 

\subsubsection{Mean free path of the ionizing photons, $R_{\rm mfp}$}
\label{subsub: mfp}
The physical size of the ionized region is governed by the distance ionizing photons propagate through the IGM, which depends on the population of the photon absorption systems where recombinations take place. To take into account this effect, we define $R_{\rm mfp}$ as the maximum horizon for the ionizing photons. 

It has been shown by \citet{Greig_2017} that this parameter is only sensitive during the later stages of reionization when the typical size of the $\HII$ regions approaches $R_{\rm mfp}$. We use a flat prior on $R_{\rm mfp} \in (5, 25)$ cMpc similar to \citet{Greig_2015}, which is consistent with the sub-grid recombination model of \citet{2014MNRAS.440.1662S}.
\begin{figure*}
    \centering
    \includegraphics[width=0.98\linewidth]{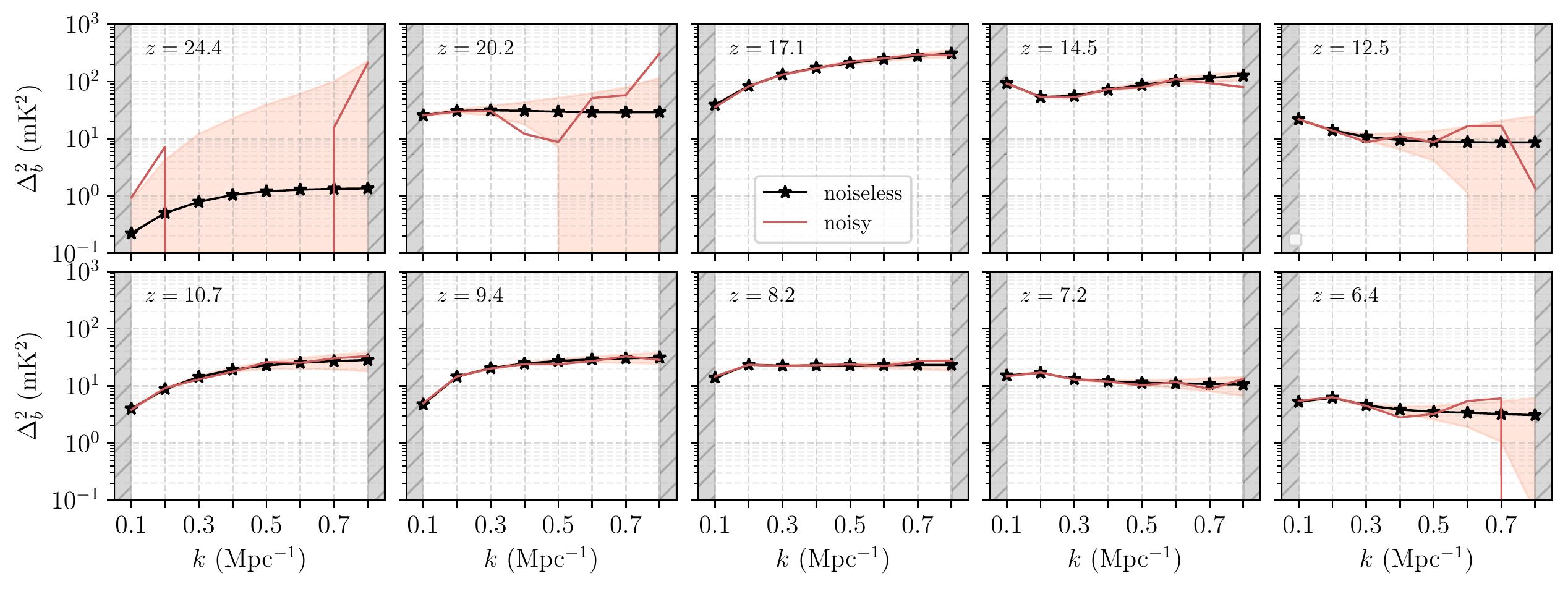}
    \caption{The cosmological (black) and noisy (orange) mock power spectrum for the {\scriptsize FAINT GALAXIES} model at different redshifts, where $k \in (0.1, 0.8)$ Mpc$^{-1}$. The shaded region represents the power spectrum uncertainty level.}
    \label{fig:mock}
\end{figure*}

\subsubsection{Integrated soft-band luminosity, $L_{\rm X < 2\, keV}/{\rm SFR}$}
The total integrated soft-band (< 2 keV) luminosity per SFR escaping the host galaxies ($L_{\rm X < 2\, keV}/{\rm SFR}$) controls the efficiency with which X-rays heat the IGM. It decides the timing and duration of the EoH in a manner similar to $\zeta$ for the EoR. 

For sufficiently large values of $L_{\rm X < 2\, keV}/{\rm SFR}$, the X-rays can also ionize the IGM at $\sim 10-20\%$ level, in addition to heating. We use a flat prior on $\log_{10}(L_{\rm X < 2\, keV}/{\rm SFR}) \in (38, 42)$. This range is motivated by population synthesis models at high redshifts \citep{2013ApJ...776L..31F} and the observations of the local population of galaxies \citep{2012MNRAS.419.2095M, 2017MNRAS.466.1019S}.

\subsubsection{X-ray energy threshold for self-absorption by the host galaxies, $E_0$}
The soft X-rays produced by the host galaxies can be absorbed by the interstellar medium, in which case they can no longer contribute to the heating of the IGM. From the simulations of high $z$ galaxies, it has been shown by \citet{Das_2017} that the attenuation of the X-ray profile can be approximated by a step function below an energy threshold $E_0$.

The small values of $E_0$ lead to very efficient and inhomogeneous heating. It has been shown by \citet{Pacucci_2014} that the amplitude of the power spectra for such softer spectral energy distributions (SEDs) is larger by up to an order of magnitude. We adopt a flat prior on $E_0 \in (0.1, 1.5)$ keV.

\subsubsection{X-ray spectral index, $\alpha_{\rm X}$} 
The spectral index governs the spectrum that emerges from the X-ray sources and depends on the dominant physical process emitting the X-ray photons. We take a flat prior on $\alpha_{\rm X} \in (-0.5, 2.5)$ similar to \citet{Greig_2017} to take into account various relevant X-ray SEDs such as HMXBs, mini-quasars, host ISM, SNe remnants.

Our simulations are performed within a [250 cMpc]$^3$ box on a  [128]$^3$ grid. The training data is composed of 20,000 power spectra samples evaluated at ten different redshifts in range (25, 6). These samples are drawn randomly from the priors. We use 80\% of the samples for training, 10\% for validation, and 10\% for the test dataset. We also vary the cosmic seed in our forward models. The impact of the size of the training data is investigated in Appendix~\ref{app:trainSubset}.

\subsection{Telescope noise profile}
\begin{table}
 \renewcommand{\arraystretch}{1.5}
 \centering
 \caption{Observation parameters for SKA1 low configuration used in this work to simulate the thermal noise.}
 \label{tab:noise_params}
 \begin{tabular}{ll}
  \hline
  Parameter & Value\\
  \hline
  $N_{\rm ant}$ & 512\\
  $\Delta \nu$ & 195.3 kHz\\
  $\Delta t$ & 10 seconds\\
  $t_{\rm obs}^{\rm day}$ & 6 hours\\
  $t_{\rm obs}^{\rm tot}$ & 1000 hours\\
  \hline
 \end{tabular}
\end{table}
\begin{figure*}
    \centering
    \includegraphics[width=0.83\linewidth]{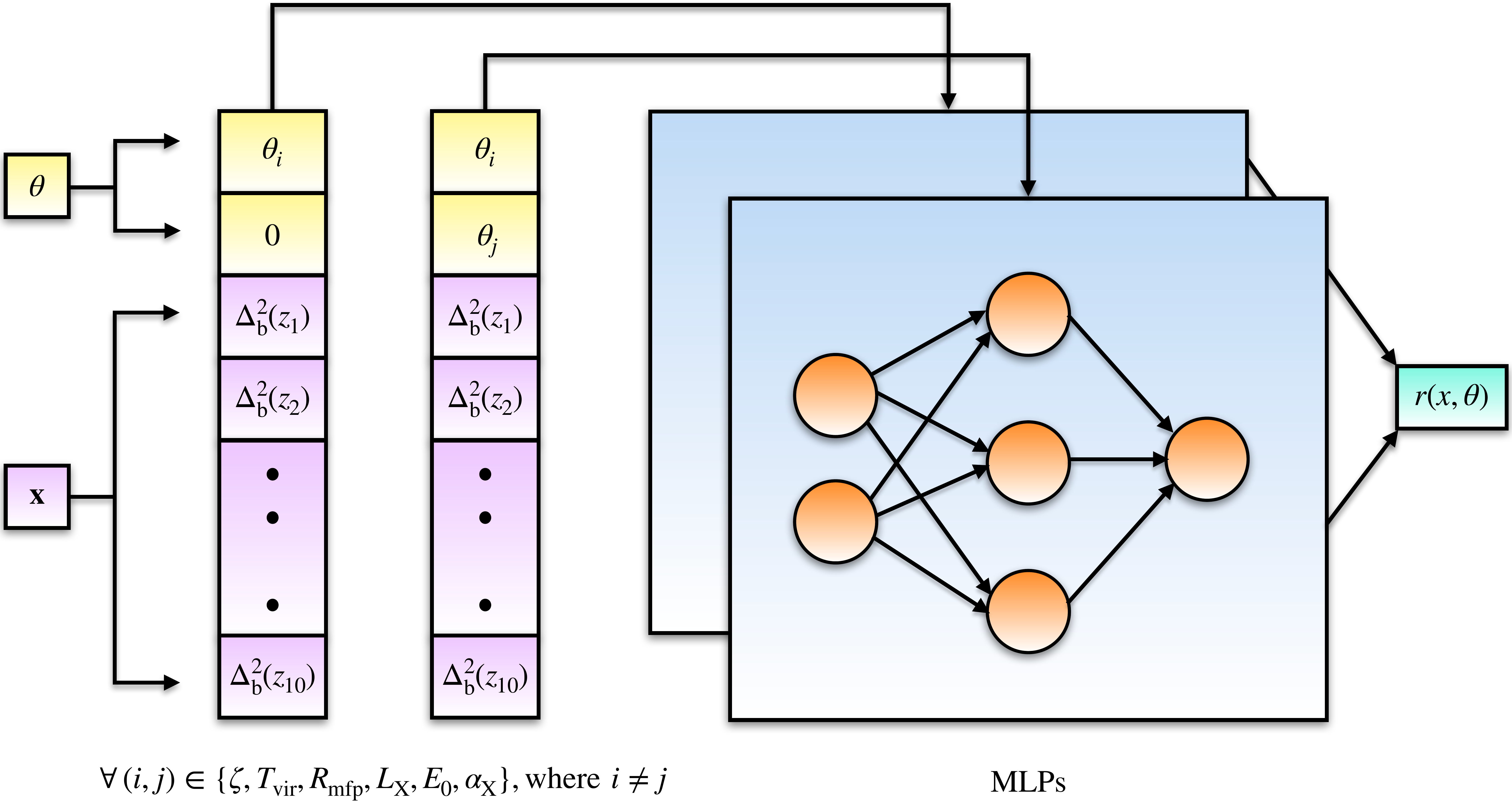}
    \caption{Illustration of the network architecture. Input data $\bm{x}$ and parameters $\bm{\theta}$ are mapped to the marginal parameter combinations. The individual ratio estimators are trained with an MLP. The outputs are estimated ratios $r(\bm{x}, \bm{\theta})$ for the marginal posteriors of interest.}
    \label{fig:net_arch} 
\end{figure*}
\begin{figure*}
    \centering
    \includegraphics[width=0.98\linewidth]{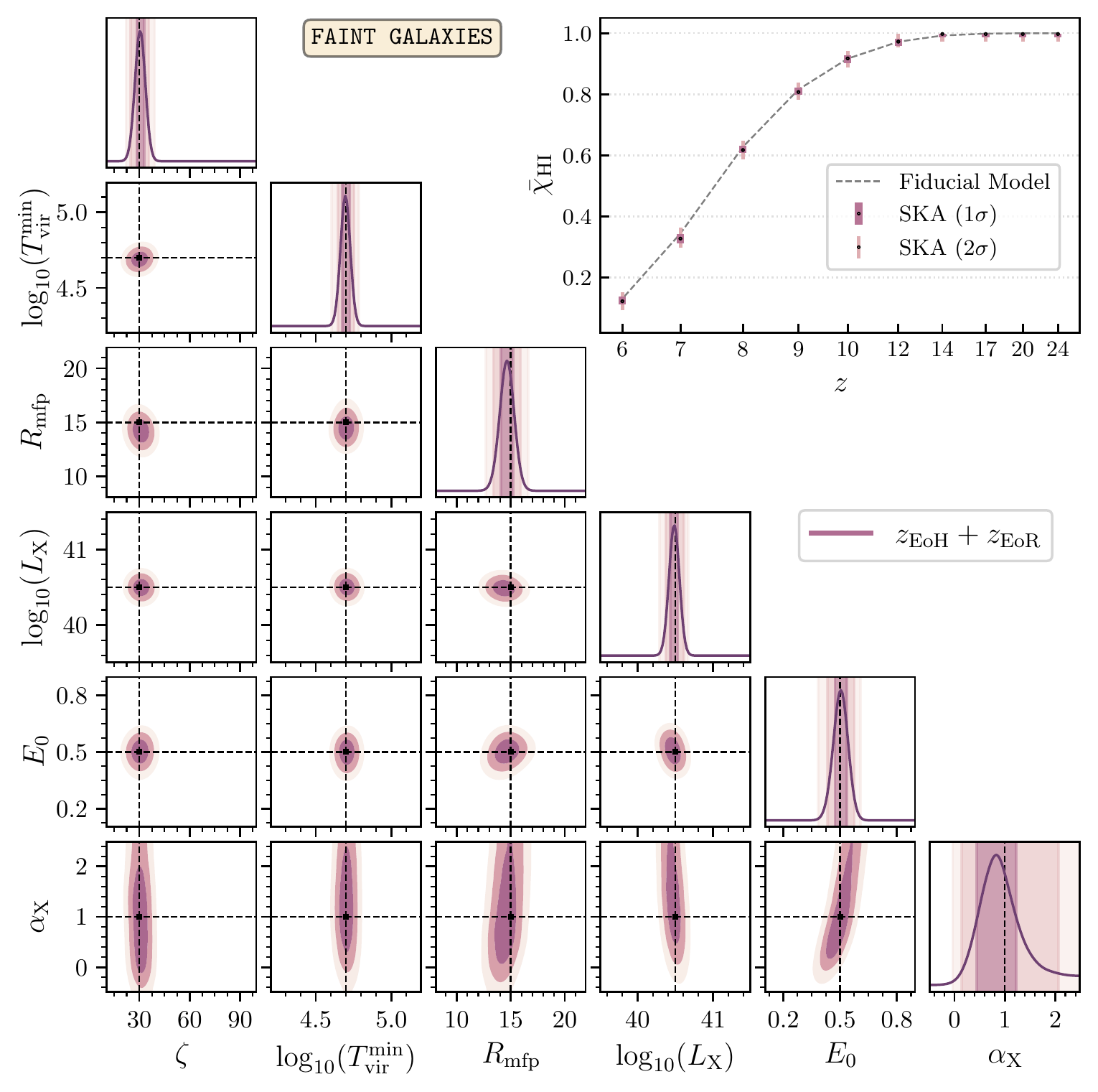}
    \caption{Recovered 1D and 2D marginals for the six-dimensional FAINT GALAXIES model assuming 1000h observation from SKA. The dashed lines denote the input parameters \{$\zeta, \log_{10}(T_{\rm vir}^{\rm min}), R_{\rm mfp}, \log_{10}(L_{\rm X}), E_0, \alpha_{\rm X}\}$ = \{30, 4.70, 15, 40.5, 0.5, 1\}. Inset: Recovered $1\sigma$ and $2\sigma$ constraints on the reionization history. The dashed line shows the evolution of $\bar{\chi}_{\HI} (z)$ for the fiducial model.}
    \label{fig:post_1D_2D}
\end{figure*}
To simulate the thermal noise, we first estimate the uv coverage for SKA1 low, assuming 1000h of observations. Thermal noise is simulated using \texttt{ps\_eor}\footnote{\url{https://gitlab.com/flomertens/ps_eor}} by creating a SEFD of 2500 Jy at the central frequency of the observation. The current configuration of SKA1-Low has 512 stations, 224 of which are placed randomly in a circular core of radius 350 m. The remaining 288 stations are distributed among 36 clusters in three spiral arms extending up to a radius of 35 km from the central core. We integrate for 10 seconds per visibility and observe for 6 hours each day with a frequency resolution of 195.3 kHz. These parameters are tabulated in Table~\ref{tab:noise_params}. This results in the thermal noise ($\sigma_{\rm therm}$).
Note that in \citet{Greig_2017}, the authors also include a 20\% modeling uncertainty on the sampled power spectra to take into account the differences with various semi-numerical and radiative transfer simulations. This can be easily incorporated without running any additional 21-cm signal simulations in our analysis. The impact of including the modeling uncertainty is discussed in Appendix~\ref{app:mod_uncert}.

\subsection{Mock observation}
To form our mock observation, we consider a model with $\{\zeta, \log_{10}(T_{\rm vir}^{\rm min}), R_{\rm mfp}, \log_{10}(L_{\rm X}), E_0, \alpha_{\rm X}\}$ = \{30, 4.70, 15, 40.5, 0.5, 1\}. It corresponds to the {\small FAINT GALAXIES} model from \citet{2016MNRAS.459.2342M, Greig_2017} in which reionization is driven by numerous sources with low ionizing efficiency. This set of parameter values results in the reionization history and the Thomson optical depth $\tau$ consistent with Planck data \citep{2016A&A...596A.108P}. The mock observation is simulated within a [500 cMpc]$^3$ box on a [256]$^3$ grid.

In Figure~\ref{fig:mock}, we show the cosmological 21-cm power spectra (black line) from our mock observation at different redshifts. The shaded region represents the 21-cm power spectrum uncertainty. We then draw a random realization from the normal distribution $\sim \mathcal{N}(0, \sigma_{\rm therm}^2 (k, z))$, and add it to the cosmological 21-cm power spectrum to form the noisy mock observation (orange line). We restrict our analysis to the $k$-modes in the range $k \in (0.1, 0.8)$ Mpc$^{-1}$ to avoid the impact of foreground contamination on large scales and thermal noise on small scales \citep{Greig_2015, Greig_2017}.

%% file: sections/results.tex
In this section, we discuss the application of \textit{swyft} to obtain the posterior probability distributions for our six-dimensional 21-cm power spectra model for the simulated mock observation and explore the constraints on different astrophysical parameters as a function of redshift in Section~\ref{subsec:post}. In Section~\ref{subsec:optim}, we show how the distribution of integration time over different redshifts can be used as a proxy to find which part of the data each model parameter is sensitive to. These examples emphasize the flexibility of our framework.

\subsection[Posterior inference with swyft]{Posterior inference with \lowercase{\textit{swyft}}}
\label{subsec:post}
\begin{table*}
 \renewcommand{\arraystretch}{1.6}
 \centering
 \caption{The inferred parameter values and the associated 16th and 84th percentiles for the posteriors from (i) $z_{\rm EoH} + z_{\rm EoR}$ shown in Figure~\ref{fig:post_1D_2D}, and (ii) $z_{\rm EoH}$ and $z_{\rm EoR}$ shown in Figure~\ref{fig:post_EoH_EoR}.}
 \label{tab:paramsRecov}
 \begin{tabular}{lllllll}
  \hline
  Model & $\zeta$ & $\log_{10}(T_{\rm vir}^{\rm min})$ & $R_{\rm mfp}$ & $\log_{10}(L_{\rm X})$ & $E_0$ & $\alpha_{\rm X}$\\
  \hline
  $z_{\rm EoH} + z_{\rm EoR}$ & $30.25_{-1.80}^{+2.70}$ & $4.70_{-0.02}^{+0.03}$ & $14.65_{-0.56}^{+0.56}$ & $40.49_{-0.06}^{+0.04}$ & $0.50_{-0.03}^{+0.03}$ & $0.84_{-0.39}^{+0.39}$\\
  \hline
  $z_{\rm EoH}$ & $22.15_{-5.40}^{+5.40}$ & $4.70_{-0.02}^{+0.03}$ & ${\centering -}$ & $40.49_{-0.06}^{+0.06}$ & $0.49_{-0.04}^{+0.03}$ & $0.68_{-0.45}^{+0.51}$\\
  $z_{\rm EoR}$ & $29.35_{-3.60}^{+2.70}$ & $4.66_{-0.05}^{+0.04}$ & $14.65_{-0.56}^{+0.56}$ & $40.47_{-0.12}^{+0.12}$ & $0.32_{-0.10}^{+0.07}$ & $-$\\
  \hline
 \end{tabular}
\end{table*}
To obtain the posterior distribution from {\small MNRE}, we first concatenate the power spectra from different redshifts into a 1D array. This is then fed as the input for the multi-layer perceptron (MLP) with three layers, each containing 256 neurons. The network is trained with a batch size of 64, and we decay the initial learning rate of $10^{-3}$ by $0.95$ after every epoch. The output of the trained network is the estimated ratios for the parameters of interest. In Figure~\ref{fig:net_arch}, we show a schematic diagram of the network architecture. Once the network is trained, the ratio estimator allows for very fast {\small MCMC} sampling from the approximate posterior.

In Figure~\ref{fig:post_1D_2D}, we present the posteriors on the astrophysical parameters obtained from \textit{swyft} for the {\small FAINT GALAXIES} model assuming 1000h observation from the SKA. The diagonal panels show the 1D marginalized posterior for each parameter, and 2D marginals are shown in the lower off-diagonal panels. The dashed lines represent the true value of the parameters. The inferred model parameters and the corresponding 16th and 84th percentiles are tabulated in  Table~\ref{tab:paramsRecov}.

Consistent with \citet{Greig_2017}, we are able to tightly constrain all our model parameters except $\alpha_{\rm X}$, which has a relatively small impact on the amplitude of the 21-cm power spectra. The small degeneracies between $\zeta - \log_{10}(T_{\rm vir}^{\rm min})$ and $E_0 - \alpha_{\rm X}$ are in agreement with \citet{Greig_2017}, and the findings of \citet{Ewall_Wice_2016} and \citet{Kern_2017}. In the top right panel of Figure~\ref{fig:post_1D_2D}, we show the $1\sigma$ and $2\sigma$ constraints on the mean neutral fraction ($\bar{\chi}_{\rm HI}$) as a function of redshift $z$, where the dashed line represents the true ionization history of the model. For this analysis, we use $\bar{\chi}_{\HI} (z)$ in place of parameters $\*\theta$ of the network architecture shown in Figure~\ref{fig:net_arch}. We find tight constraints on the ionization history from the SKA.

\begin{figure*}
    \centering
    \includegraphics[width=0.98\linewidth]{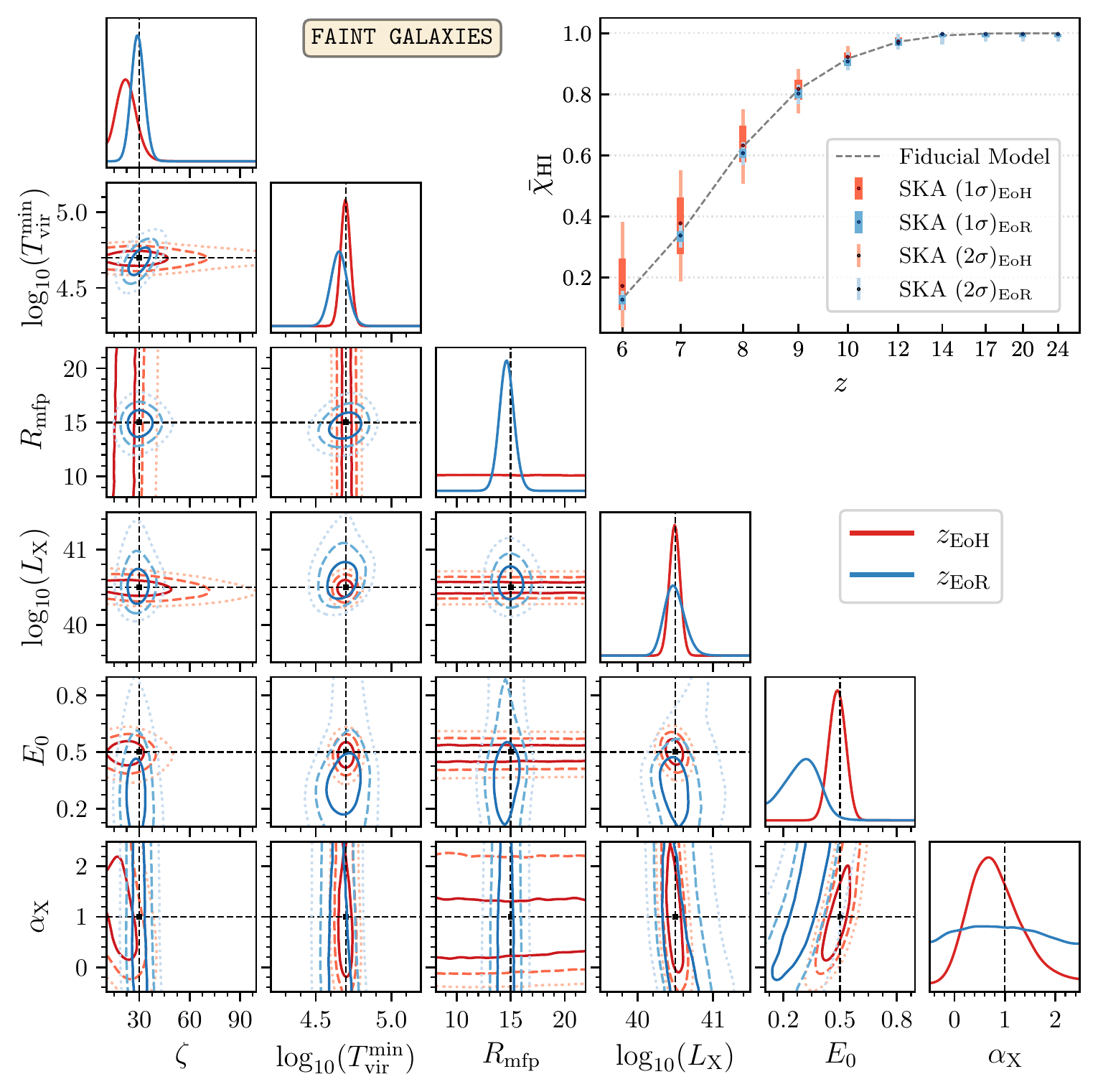}
    \caption{Recovered 1D and 2D marginals from $z_{\rm EoH}$ (red) and $z_{\rm EoR}$ (blue) for the six-dimensional FAINT GALAXIES model assuming 1000h observation with SKA. The dashed lines denote the input parameters \{$\zeta, \log_{10}(T_{\rm vir}^{\rm min}), R_{\rm mfp}, \log_{10}(L_{\rm X}), E_0, \alpha_{\rm X}\}$ = \{30, 4.70, 15, 40.5, 0.5, 1\}. The inset plot shows the recovered reionization history from the 21-cm power spectra during $z_{\rm EoH}$ (red) and $z_{\rm EoR}$ (blue).}
    \label{fig:post_EoH_EoR}
\end{figure*}
\begin{figure*}
    \centering
    \includegraphics[width=\linewidth]{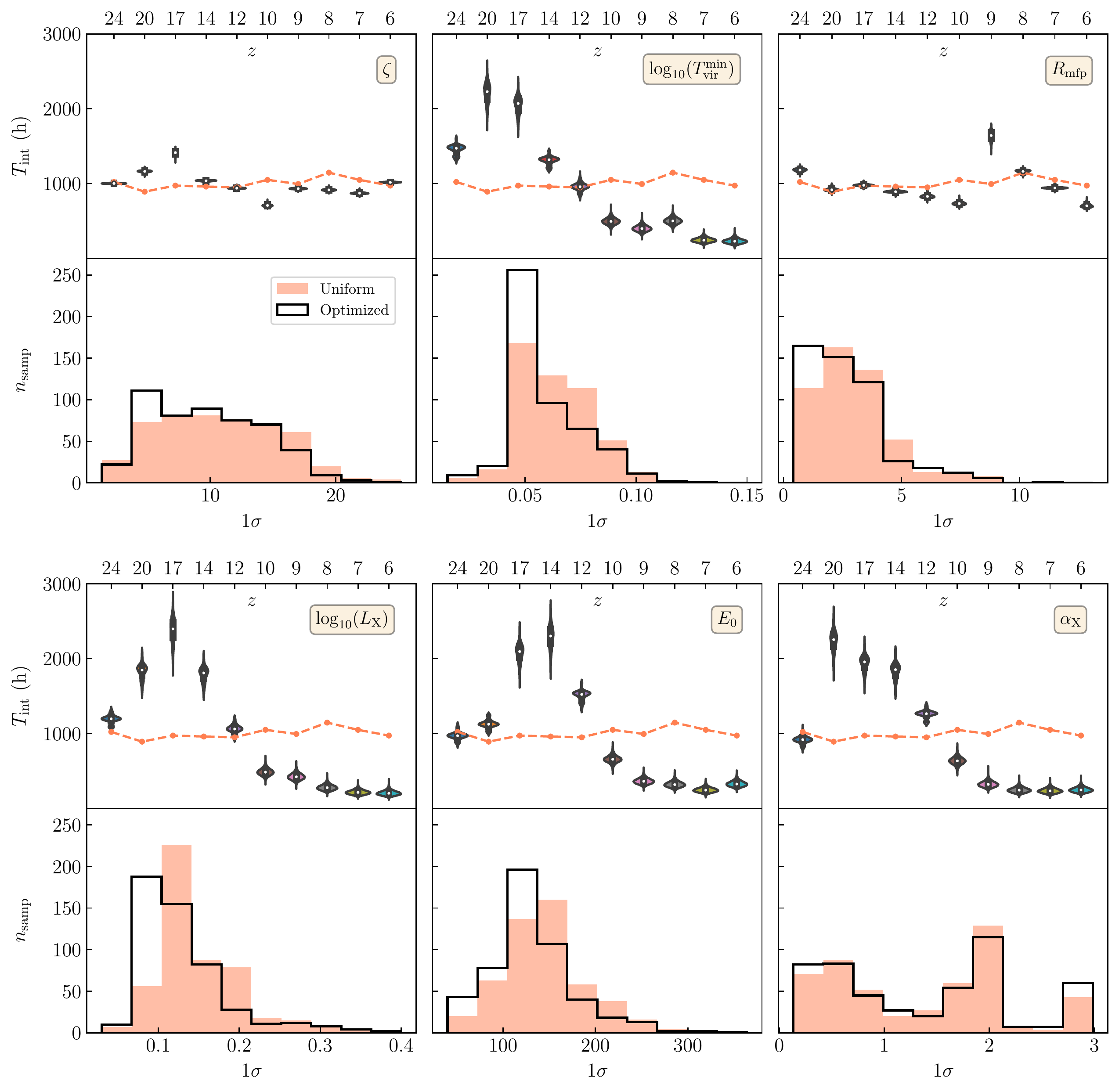}
    \caption{Result of integration time optimization as a proxy for information content for $\zeta$, $\log_{10}(T_{\rm vir}^{\rm min})$, $R_{\rm mfp}$, $\log_{10}(L_{\rm X})$, $E_0$, $\alpha_{\rm X}$. \textbf{Top panel:} Violins represent the optimized time distribution, and the orange line shows the initial time distribution. \textbf{Bottom panel:} The histogram of 1$\sigma$ uncertainty interval on the posterior distribution of each parameter for 500 different mock observations drawn randomly from the test dataset assuming a uniform (orange) and optimized (black) time distribution.}
    \label{fig:timeOptim}
\end{figure*}

Next, we investigate the sensitivity of our model parameters during different redshifts. We perform the parameter inference by dividing the entire redshift range into two bins: (i) $z_{\rm EoH} \in (25, 12)$, which corresponds to the X-ray heating, and (ii) $z_{\rm EoR} \in (11, 6)$ that corresponds to reionization. Note that this analysis does not require any extra 21-cm power spectra simulations. The same training data can be re-used with a minimal change in the network's architecture, which is not possible for an {\small MCMC} analysis. In this case, the MLP takes the power spectra from $z_{\rm EoH}$ (or $z_{\rm EoR}$) as the input and estimates the ratios for the parameters of interest.

In Figure~\ref{fig:post_EoH_EoR}, we show the resulting 1D and 2D marginal posteriors from $z_{\rm EoH}$ (red) and $z_{\rm EoR}$ (blue). The inferred parameters and the corresponding 16th and 84th percentiles are tabulated in Table~\ref{tab:paramsRecov}. We find that $\zeta$, $\log_{10}(T_{\rm vir}^{\rm min})$ and $R_{\rm mfp}$ are well constrained with the 21-cm power spectra from $z_{\rm EoR}$, which is expected as these parameters play a significant role during reionization. On the other hand, $\log_{10}(T_{\rm vir}^{\rm min})$, $\log_{10}(L_{\rm X})$ and $E_0$ are constrained with the power spectra from $z_{\rm EoH}$. Note that the minimum virial temperature of a halo to host the star-forming galaxies, $T_{\rm vir}^{\rm min}$ can be well constrained with either redshift bin because, within \texttt{21cmFAST}, the galaxies that host the ionizing sources are the same galaxies that are responsible for X-ray heating. So, this parameter impacts both the EoH and EoR.

In the top right panel of the Figure~\ref{fig:post_EoH_EoR}, we show the constraints on reionization history from the 21-cm power spectra during $z_{\rm EoH}$ (red) and $z_{\rm EoR}$ (blue). We find that from the 21-cm power spectra at $z_{\rm EoH}$, we can constrain the neutral fraction reasonably well at $z\geq 12$ ($z_{\rm EoH}$), but it does not provide tight constraints during the intermediate and late stages of reionization. However, with the 21-cm power spectra at $z_{\rm EoR}$, we can infer the entire reionization history of the {\small FAINT GALAXIES} model. The constraints from the 21-cm power spectra at $z_{\rm EoR}$ on the neutral fraction at $z_{\rm EoH}$ comes from the fact that throughout our models, the neutral fraction $\bar{\chi}_\HI \approx 1$ at $z \geq 12$.

\subsection{Information from different redshifts}
\label{subsec:optim}
In order to study the information content from different redshifts, we consider a toy scenario where we use the distribution of integration time over different redshifts as a proxy to find which part of the data each parameter is most sensitive to. So far, in our analysis, we considered the distribution of integration time to be uniform over redshifts. However, for a fixed total integration time, this distribution can be optimized since the thermal noise level at redshift $z$ depends on 
the integration time $t_z$ allocated for that redshift.

The optimization is achieved via gradient descent by maximizing the information the network learns about any given parameter from different redshifts. For a fixed total integration time $T_{\rm tot}$, we optimize the integration time for each redshift $t_{z}$, such that $T_{\rm tot}= \sum t_z$. We parameterize $t_{z}$ as 
\begin{equation}
    t_{z} = \left(T_{\rm tot}\times\verb|softmax|(\*v)\right)_{z}\,,
\end{equation}
where $\*v$ is a vector that corresponds to the number of redshift bins. Larger components in $\*v$ correspond to more integration time for that redshift bin. Next, we consider $\*v$ to be one of the network parameters that is optimized during the training. As we train the classifier (MLP) to learn the 1D posterior for any given parameter, at the same time, it learns the optimal way of distributing the integration time for that parameter. We further obtain the uncertainties on the optimal time distribution through Monte Carlo Dropout ({\small MCD}) \citep{https://doi.org/10.48550/arxiv.1506.02142}.

In Figure~\ref{fig:timeOptim}, we show the results from this information content analysis. For each parameter, the top panel shows the uniform time distribution (orange dashed line) and the optimized time distribution (violins), where the uncertainties follow from the {\small MCD}. The bottom panel shows the histogram of the $1\sigma$ uncertainty interval on the posterior distribution of each parameter from 500 different mock observations drawn randomly from the test dataset assuming the uniform (orange) and optimized (black) time distribution.

We find that the parameters $\log_{10}(T_{\rm vir}^{\rm min})$, $\log_{10}(L_{\rm X})$, $E_0$ and $\alpha_{\rm X}$ are assigned a larger integration time at high redshifts $z \geq 12$ after the optimization of the network. This implies that the information for these parameters is contained at high redshifts. These findings are consistent with the posteriors from $z_{\rm EoH}$ and $z_{\rm EoR}$ shown in Figure~\ref{fig:post_EoH_EoR}. On the contrary, for the mean free path $R_{\rm mfp}$, the network allocates large integration time at redshift $z$ = 9. This is also in agreement with the analysis in Figure~\ref{fig:post_EoH_EoR}, where we found a flat posterior on $R_{\rm mfp}$ from $z_{\rm EoH}$, and the constraints only came from $z_{\rm EoR}$. For each parameter, the histogram of the $1\sigma$ uncertainty interval from the optimized integration time distribution tends towards lower $1\sigma$ uncertainty on the posterior distribution, which indicates that the network learns more information about a given parameter from the optimized time distribution in comparison to the uniform time distribution.

%% file: sections/summary.tex
In this paper, we performed Simulation-Based Inference through a {\small MNRE} algorithm, \textit{swyft}, to constrain the astrophysical parameters that govern the X-ray heating and reionization during the CD-EoR. We used \texttt{21cmFAST} to model the 21-cm power spectra during CD-EoR with a six-dimensional astrophysical parameter space. We showed that this framework is significantly more efficient as it directly learns the marginal posteriors of interest through neural networks than the conventional likelihood-based methods such as {\small MCMC}, which samples the full joint posterior. 

With the training data composed of 20,000 21-cm power spectra simulations and the expected thermal noise level from the SKA, we were able to constrain the parameters of our model. The 1D and 2D marginal posteriors obtained through {\small MNRE} look consistent with the earlier studies performed with an {\small MCMC}, which required an order of magnitude more samples to converge. We further checked the statistical consistency of the trained network by evaluating the nominal and empirical expected coverage probabilities.

Within \textit{swyft}, generating the training dataset and {\small MNRE} are two independent processes. This feature gives us the flexibility to reuse the simulations and utilize the same training dataset for various applications. To demonstrate this aspect of \textit{swyft}, we investigated the sensitivity of different parameters over two different redshift ranges that correspond to the EoH ($z_{\rm EoH}$) and EoR ($z_{\rm EoR}$). We obtained the posterior probability distribution on the model parameters from $z_{\rm EoH}$ and $z_{\rm EoR}$ at no extra cost of 21-cm power spectra simulation. An {\small MCMC} analysis in this scenario would otherwise require a new chain, and the simulations can not be used efficiently.

We further studied the information content for each parameter from different redshifts by considering a toy scenario where we consider the distribution of the integration time to be part of the network parameters, which is optimized during the training. We found the optimized time distribution to be consistent with the posterior probability distribution of model parameters from $z_{\rm EoH}$ and $z_{\rm EoR}$. This could be used as an indicator of the possible degeneracies for more complex astrophysical 21-cm signal models without running additional simulations. This establishes that with such efficient and scalable inference techniques, one can increase the complexity of the 21-cm model even further, which could otherwise be impractical for the likelihood-based approaches.

While our analysis has shown that {\small MNRE} is a powerful framework to analyse the 21-cm power spectrum,  in reality, the 21-cm signal during CD-EoR is highly non-Gaussian \citep{Shimabukuro_2016, Majumdar_2018, Watkinson_2018}, so the 21-cm power spectrum is probably not the most optimal summary statistics to use for parameter inference. In future work, we plan to explore the higher-order summary statistics such as the 21-cm bispectrum \citep{Tiwari_2022}, the morphology of the ionized regions \citep{Gazagnes_2021, Kapahtia_2021} and convolutional neural networks on the 21-cm tomographic images \citep{Gillet_2019, Zhao_2022_3D} for parameter inference through {\small MNRE}.